\documentclass[aps,prb,preprint,groupedaddress,showpacs,amsmath,amssymb]{revtex4}

\usepackage{graphicx}
\usepackage{dcolumn}
\usepackage{bm}

\begin{document}

\title{Effect of pressure on the structural phase transition and superconductivity in (Ba$_{1-x}$K$_x$)Fe$_2$As$_2$ ($x = 0$ and $0.45$) and SrFe$_2$As$_2$ single crystals.}

\author{M. S. Torikachvili}
\affiliation{Department of Physics, San Diego State University, San Diego, CA 92182-1233}
\author{N. Ni}
\author{S. L. Bud'ko}
\author{P. C. Canfield}
\affiliation{Ames Laboratory US DOE and Department of Physics and Astronomy, Iowa State University, Ames, Iowa 50011}

\date{\today}

\begin{abstract}
The effects of pressure up to $\sim 20$ kbar, on the structural phase transition of SrFe$_2$As$_2$ and lightly Sn-doped BaFe$_2$As$_2$, as well as on the superconducting transition temperature and upper critical field of (Ba$_{0.55}$K$_{0.45}$)Fe$_2$As$_2$ single crystals have been studied. All the  transitions temperatures decrease with pressure in an almost linear fashion. Under pressure, the upper critical field curve, $H_{c2}(T)$, for (Ba$_{0.55}$K$_{0.45}$)Fe$_2$As$_2$ shifts down in temperature to follow the zero field $T_c$ with very little change in slope. Apparent similarity in the temperature - pressure phase diagrams for three AFe$_2$As$_2$ (A = Ba, Sr, Ca) parent compounds is noted.
\end{abstract}


\maketitle

\section{Introduction}
The recent reports of the discovery of elevated temperature superconductivity in LaFeAs(O$_{1-x}$F$_{x}$) \cite{jap_dis} followed by an almost two-fold increase in the superconducting transition temperature by application of pressure \cite{jap_pre} or by substitution of heavier rare earths \cite{fst_sm} has brought a lot of attention to materials with structures containing Fe-As layers. Within a few months, superconductivity below $T_c = 38$ K was discovered in the structurally related, non-oxide material, K-doped BaFe$_2$As$_2$ \cite{sec_ger} and single crystals of the parent, non-superconducting compound, BaFe$_2$As$_2$, as well as superconducting (Ba$_{0.55}$K$_{0.45}$)Fe$_2$As$_2$ were synthesized. \cite{ni_ba,ba_cn} Although the parent compound appears to be sensitive to small, $\sim 1$\%, of Sn doping, \cite{ni_ba,ba_cn,fst_ger} a small amount of Sn does not seem to be detrimental for superconductivity in the K-doped compound (light doping of Sn is an unintended consequence of the crystal growth process). \cite{sec_ger,ni_ba} Soon after that single crystals of pure and K-doped SrFe$_2$As$_2$ were synthesized. \cite{yan,sr_cn} SrFe$_2$As$_2$ bears similarity to BaFe$_2$As$_2$ in having structural (antiferromagnetic) phase transition at $\sim 200$ K and  exhibiting superconductivity at elevated temperatures upon K-doping. Bearing in mind the significant effect of pressure on the structural phase transition and superconductivity in the related, RFeAs(O$_{1-x}$F$_{x}$) (R = rare earth), family of materials, in this work we study the effect of hydrostatic pressure on parent, lightly Sn-doped, BaFe$_2$As$_2$,  superconducting (Ba$_{0.55}$K$_{0.45}$)Fe$_2$As$_2$ as well as parent SrFe$_2$As$_2$ single crystals.

\section{Experimental details}

Single crystals of BaFe$_2$As$_2$, (Ba$_{0.55}$K$_{0.45}$)Fe$_2$As$_2$, and SrFe$_2$As$_2$  were grown out of a Sn flux, using conventional high temperature solution growth techniques. \cite{can_fi}  The details of the growth as well as thermodynamic and transport properties of these crystals are described in Refs. [\onlinecite{ni_ba,yan}. At ambient pressure the structural phase transition in BaFe$_2$As$_2$ is marked by a rapid increase of in-plane resistivity, whereas in SrFe$_2$As$_2$ in-plane resistivity abruptly decreases below such transition.
The pressure dependencies of the structural phase transition temperature, $T_0$, the superconducting phase transition temperature, $T_c$, and the upper critical field, $H_{c2}$, were determined from the temperature-dependent in-plane resistance. Pressure was generated in a Teflon cup filled with either a 60:40 mixture of n-pentane and light mineral oil (BaFe$_2$As$_2$) or Fluorinert FC-75 ((Ba$_{0.55}$K$_{0.45}$)Fe$_2$As$_2$) inserted in a 22 mm outer diameter, non-magnetic, piston-cylinder-type, Be-Cu pressure cell with a core made of NiCrAl (40 KhNYu-VI) alloy. The pressure at room temperature was monitored by a manganin, resistive pressure gauge. At low temperatures the pressure value was determined from the superconducting transition temperature of pure lead \cite{eil81a}. Low temperature pressure values will be used throughout the text (except for part of the data for SrFe$_2$As$_2$ where either room temperature values of pressure or interpolation for intermediate temperatures will be used) as the pressure remains almost constant in similar geometry cells on cooling below $\sim 100$ K. \cite{joe}  The temperature and magnetic field environment for the pressure cell was provided by a Quantum Design Physical Property Measurement System (PPMS-9) instrument. An additional Cernox sensor, attached to the body of the cell, served to determine the temperature of the sample for these measurements. The cooling rate was below 0.5 K/min, the temperature lag between the Cernox on the body of the cell and the system thermometer was $< 0.5$ K at high temperatures and 0.1 K or less below $\sim 70$ K.

\section{Results}

\begin{figure}
\begin{center}
\includegraphics[angle=0,width=150mm]{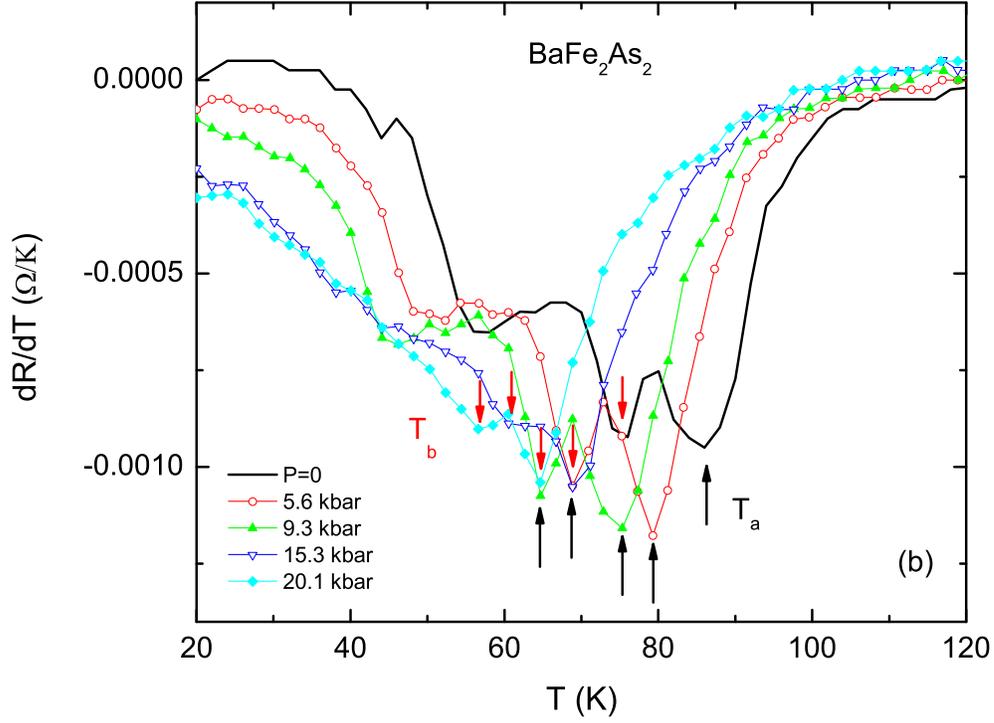}
\end{center}
\caption{(Color online) (a)Pressure dependence of in-plane resistance of BaFe$_2$As$_2$. Inset: pressure dependent transition temperatures, determined as shown in panel (b), the lines are from linear fits. (b)Derivatives, $dR/dT$, at different pressures in the transition region. Arrows show two definitions of the transition temperature.}\label{F1}
\end{figure}

Figure \ref{F1}(a) shows the temperature dependent resistance of BaFe$_2$As$_2$ at different pressures. Above $\sim 10$ kbar the $\rho(T)$ curves shift down with pressure. Although the feature in resistivity corresponding to the structural phase transition is somewhat broad, its pressure dependence can be monitored by following the pressure evolution of the minima in the derivative, $dR/dT$ (Fig. \ref{F1}(b)). It is noteworthy, that at least two minima, can be seen in $dR/dT$ (marked with up- ($T_a$) and down- ($T_b$) arrows in Fig. \ref{F1}(b)). Both minima shift to lower temperatures (inset to Fig. \ref{F1}(a)) under pressure with similar pressure derivatives, $dT_a/dP = -1.04 \pm 0.04$ K/kbar, $dT_b/dP = -0.89 \pm 0.05$ K/kbar. Using a linear extrapolation of these data, the structural phase transition can be expected to be suppressed by $\sim 80$ kbar. This is most likely an upper limit given the possibility of super-linear suppression at higher pressures.
\\

\begin{figure}
\begin{center}
\includegraphics[angle=0,width=150mm]{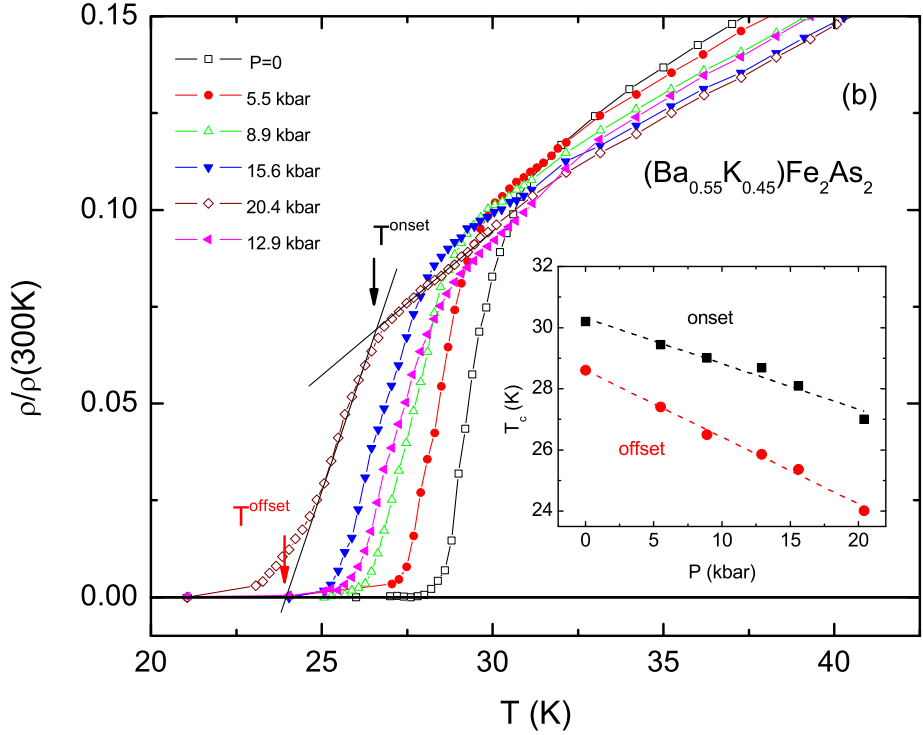}
\end{center}
\caption{(Color online) (a) Temperature dependent in-plane resistance of (Ba$_{0.55}$K$_{0.45}$)Fe$_2$As$_2$ under pressure (pressure values in the legend are in the order of runs). Dashed line - 20.4 kbar data scaled in a way that brings the last, 12.9 kbar, data between 8.9 kbar and 15.6 kbar runs. Inset: normalized temperature dependent resistivity as a function of pressure. (b) Resistive superconducting transition in (Ba$_{0.55}$K$_{0.45}$)Fe$_2$As$_2$ (pressure values in the legend are in the order of runs). Inset: superconducting transition temperatures (defined as onset and offset of resistive transition) as a function of pressure. The lines are from linear fits.}\label{F2}
\end{figure}

For superconducting (Ba$_{0.55}$K$_{0.45}$)Fe$_2$As$_2$, the normal state resistivity decreases under pressure up to $\sim 20$ kbar as shown in Fig. \ref{F2}(a).  However the normalized resistivity, $\rho(T)/\rho(300K)$ does not change significantly (Fig. \ref{F2}(a), inset). (Note that apparently the sample has developed a small crack after the 4-th pressure run, 15.6 kbar, that caused an increase in the measured resistance for the next two runs, however the 20.4 kbar run yields  consistent values if scaled with the last, 12.9 kbar, curve, as shown in Fig. \ref{F2}(a)) The superconducting transition temperature decreases under pressure with some (reversible) broadening of the resistive transition (Fig. \ref{F2}(b)). For different criteria in the determination of $T_c$, the pressure derivatives are  $dT_c^{onset}/dP = -0.15 \pm 0.01$ K/kbar, $dT_c^{offset}/dP = -0.21 \pm 0.01$ K/kbar (Fig. \ref{F2}(b)).

\begin{figure}
\begin{center}
\includegraphics[angle=0,width=150mm]{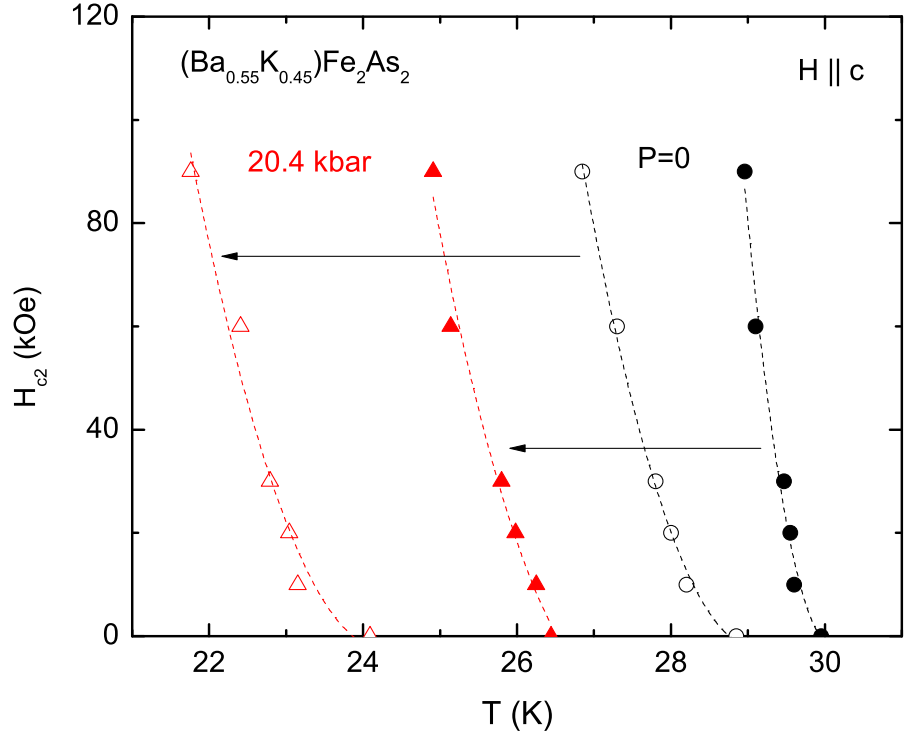}
\end{center}
\caption{(Color online) Upper critical fields, $H_{c2}(T)$ measured up to 90 kOe at zero and maximum pressure as defined from onset (filled symbols) and offset (open symbols) of resistive transitions. The lines are guides to the eye.}\label{F3}
\end{figure}

The upper critical field in (Ba$_{0.55}$K$_{0.45}$)Fe$_2$As$_2$ is expected to be extremely high, on the order of 1,000 kOe \cite{ni_ba,chuck,sin}. Our measurements, up to 90 kOe, can probe only a small section of the $H_{c2}(T)$ curve, close to zero-field $T_c$. Under pressure, the $H_{c2}(T)$ appear to shift (Fig \ref{F3}) following the shift of $T_{c0}$ without changes in the slope or curvature.
\\

\begin{figure}
\begin{center}
\includegraphics[angle=0,width=150mm]{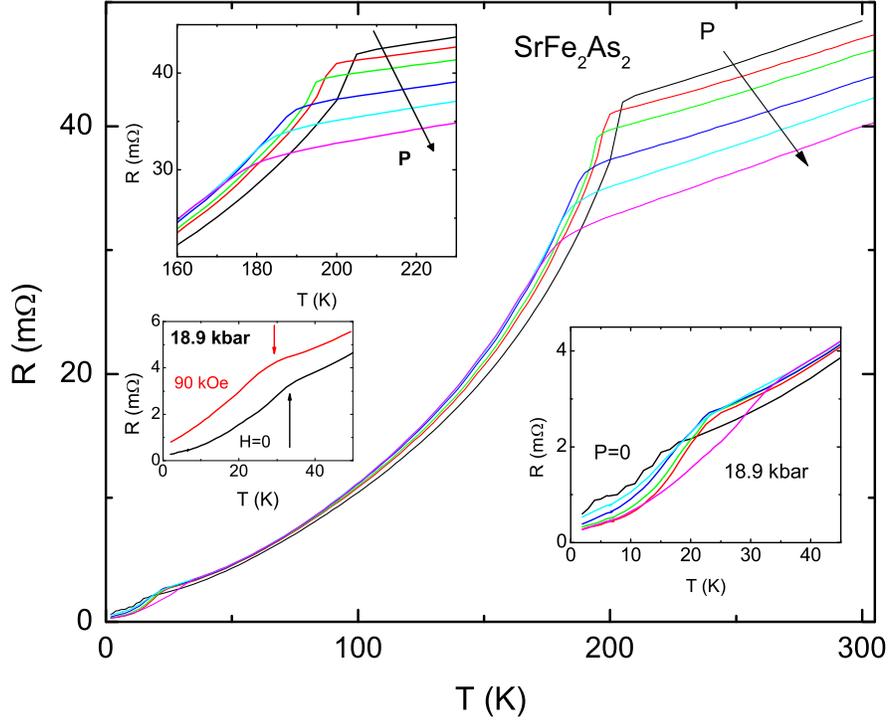}
\end{center}
\caption{(Color online) Temperature-dependent in-plane resistance of SrFe$_2$As$_2$ measured at $P = 0, 3.1, 6.4, 12.1, 15.7$ and 18.9 kbar (pressure values at low temperatures). Arrow indicates the direction of the pressure increase. Upper left inset: enlarged region of the $R(T)$ curves near the structural (antiferromagnetic) phase transition. Arrow indicates the direction of the pressure increase. Lower right inset: enlarged low temperature part of the $R(T)$ curves in the region of the $T^*$ (see text). Lower left inset: low temperature part of the $R(T)$ curves taken at $P = 18.9$ kbar for $H = 0$ and 90 kOe. Vertical arrows mark $T^*$.}\label{F4}
\end{figure}

\begin{figure}
\begin{center}
\includegraphics[angle=0,width=120mm]{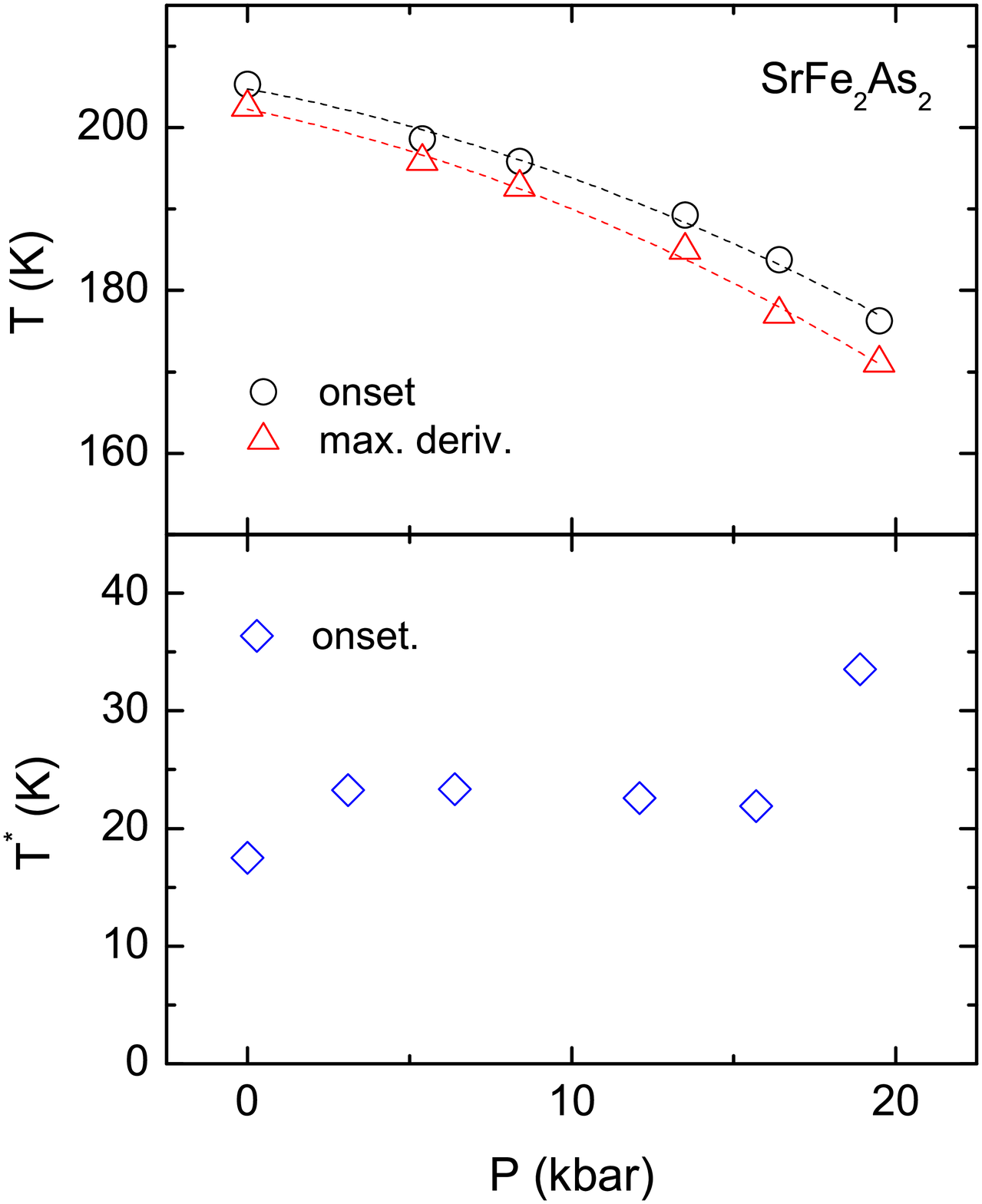}
\end{center}
\caption{(Color online) Pressure dependence of the resistance at 300 K (top panel), structural (antiferromagnetic) phase transition (middle panel) and $T^*$ - low temperature anomaly (bottom panel). Room temperature pressure values were used for $R_{300K}(P)$, interpolation consistent with Ref. [\onlinecite{joe}] were used for the middle panel and low temperature values of pressure for $T^*(P)$. Two criteria, onset and maximum of the $dR/dT$ derivatives were used for the structural (antiferromagnetic) and $T^*$ transitions. For the latter double symbols indicate that two close lying maxima in $dR/dT$ were observed. Dashed lines are second order polynomial fits.}\label{F5}
\end{figure}

Temperature dependent resistance of SrFe$_2$As$_2$ at different pressures is shown in Fig. \ref{F4}. Applied pressure noticeably lowers the high temperature (tetragonal phase) resistance. The tetragonal to orthorhombic, structural phase transition temperature is suppressed under pressure. Low temperature part of the $R(T)$ curves has an additional feature (Fig. \ref{F4}, lower right inset). This feature, $T^*$, is marked by a rapid decrease in resistance, is seen in $P = 0$ data with some ambiguity, but is unambiguously present at elevated pressures. This feature is only slightly field-dependent: for $P = 18.9$ kbar data $T^*$ shifts down by $\approx 4$ K ($< 15\%$) in 90 kOe magnetic field ($H \perp c$). Fig. \ref{F5} summarizes the pressure dependencies of the resistance at 300 K and two salient temperatures for SrFe$_2$As$_2$. $R_{300K}$ decreases non-linearly by $\sim 18\%$ at 20 kbar. The $R_{300K}(P)$ extrapolates to zero at rather moderate pressure of $\sim 50$ kbar that suggests that this initial, low pressure, $R_{300K}(P)$ behavior will probably change either gradually or through some kind of phase transition at $P > 20$ kbar. The decrease of the structural (antiferromagnetic) transition temperature under pressure is somewhat non-linear as well, by 20 kbar the transition temperature is $\sim 86\%$ of its $P = 0$ value, and a gross extrapolation of it's pressure dependence suggests that the structural transition will be suppressed by $\sim 80$ kbar. $T^*$ depends on pressure non-monotonically and apparently just starts to increase near the limit of the present measurements.

\section{Discussion and Summary}

\begin{figure}
\begin{center}
\includegraphics[angle=0,width=150mm]{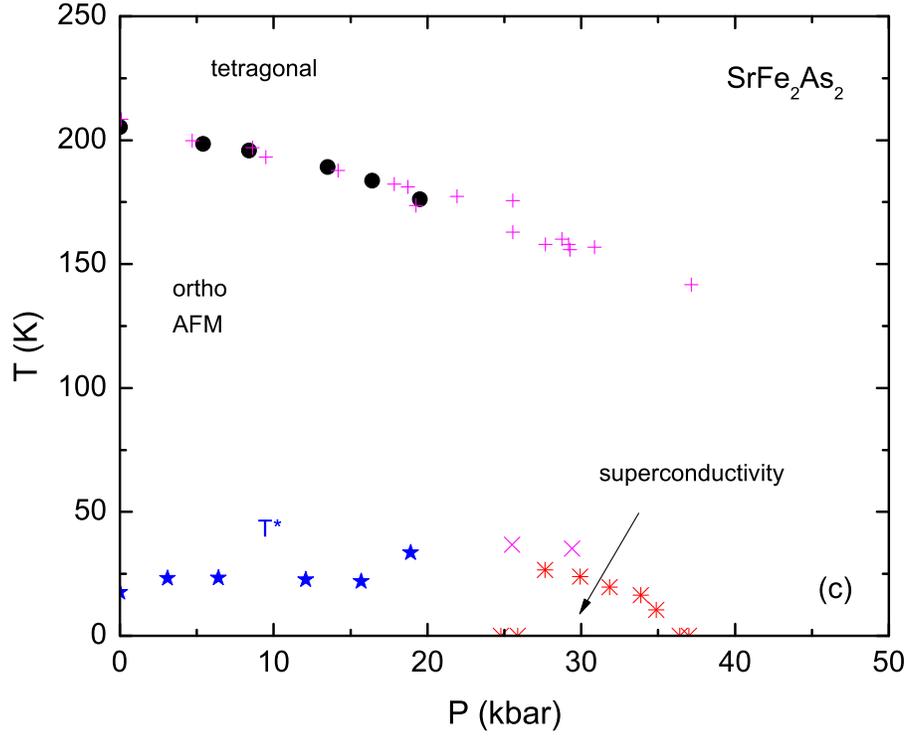}
\end{center}
\caption{(Color online) Schematic $P - T$ phase diagrams for (a) CaFe$_2$As$_2$, (b) BaFe$_2$As$_2$, (c) CaFe$_2$As$_2$. Data in panel (a) are from Ref. [\onlinecite{milt}]. Panel (b) combines data from this work ($\bullet$), Ref. [\onlinecite{bri}] ($\ast$) and shifted by 40 kbar data for (Ba$_{0.55}$K$_{0.45}$)Fe$_2$As$_2$ ($\star$) from this work. Panel (c) presents data from this work ($\bullet, \star$), Ref. [\onlinecite{bri}] ($\ast$) and Ref. [\onlinecite{ger}] ($+, \times$). }\label{F6}
\end{figure}

Of three parent AFe$_2$As$_2$ (A = Ba, Sr, Ca) compounds, the pressure - temperature phase diagram of CaFe$_2$As$_2$ is at this point studied in more details. At room temperature and ambient pressure the material has tetragonal structure and is not magnetically ordered \cite{ni_ca}. On cooling at ambient pressure, a structural, tetragonal to orthorhombic, first order, phase transition, coincident with a transition to long range antiferromagnetically ordered phase, occurs at $\sim 170$ K. \cite{ni_ca,gold}  On application of very moderate ($\sim 3$ kbar) pressure, superconductivity below $\sim 12$ K, as evidenced by zero resistivity and enhanced diamagnetic signal, \cite{milt,milt1} is observed. On further increase of pressure, superconducting transition temperature slightly increases, then decreases, and vanishes above $\sim 9$ kbar. Starting from $P > 5$ kbar, a new, high-temperature, highly hysteretic, feature in resistivity with its critical temperature rapidly rising under pressure, is observed. \cite{milt} Recent neutron diffraction studies under pressure \cite{kre} revealed that this latter transition is from high temperature tetragonal to low temperature "collapsed" tetragonal phase with smaller unit cell volume and no long range magnetic order. In the same study a low temperature, almost vertical in $P - T$ coordinates, phase boundary between the orthorhombic antiferromagnetic and collapsed tetragonal phases was suggested. A schematic $P _ T$ phase diagram based on the Refs. \onlinecite{ni_ca,gold,milt,kre} is shown in Fig. \ref{F6}(a). 

Following the discovery of superconductivity under pressure in CeFe$_2$As$_2$, \cite{milt} pressure-induced superconductivity above $\sim 25$ kbar was reported in BaFe$_2$As$_2$ and SrFe$_2$As$_2$ based on magnetic susceptibility measurements. \cite{bri} Fig. \ref{F6}(b) presents a schematic $P - T$ phase diagram for BaFe$_2$As$_2$ that combines results of Ref. [\onlinecite{bri}] and the present work. On a schematic level, this phase diagram is very similar to the one for CeFe$_2$As$_2$ (Fig. \ref{F6}(a)). At this time we are unaware of structural or electrical transport data above $\sim 20$ kbar, these data, when available, will shed light on the degree of universality of the $P - T$ phase diagrams in AFe$_2$As$_2$ (A = Ba, Sr, Ca) compounds. For comparison, the $T_c(P^*)$ data for (Ba$_{0.55}$K$_{0.45}$)Fe$_2$As$_2$ are added to the $P - T$ phase diagram (Fig. \ref{F6}(b)). These data are plotted as a function of $P^* = P + 40$ kbar, shifted along the $x$-axis so that the ambient pressure $T_c$ for (Ba$_{0.55}$K$_{0.45}$)Fe$_2$As$_2$  is close to a $T_c$ value of pure BaFe$_2$As$_2$ on a superconducting dome. This comparison plot shows that the $T_c$ of the pure sample is much more sensitive to pressure than that of the doped sample. It suggests that pressure and doping are not strictly equivalent, and a parameter more complex than pressure is needed if one attempts to a scaling of superconducting transition temperatures in pure and doped BaFe$_2$As$_2$.

Results of the current work and the data from the references [\onlinecite{bri,ger}] for SrFe$_2$As$_2$ are summarized in a phase diagram in Fig. \ref{F6}(c). Structural (antiferromagnetic) phase transition temperature decreases with pressure in agreement with the literature data \cite{ger}. Structural or electrical transport data in a wide temperature range at pressures above $\sim 35$ kbar are required to confirm or refute an existence of high pressure, tetragonal to "collapsed" tetragonal structural phase transition in SrFe$_2$As$_2$. The low temperature part of the $P - T$ phase diagram for this material appears to be more complex than for two other parent compounds in the  AFe$_2$As$_2$ (A = Ba, Sr, Ca)  family. The feature in resistivity that is very similar in shape to the $T^*$ one in this work was associated \cite{ger} with a superconducting transition, that is very tempting, having in mind the nearness of these points in $P - T$ phase diagram to the superconducting "dome" suggested by susceptibility measurements \cite{bri} (Fig. \ref{F6}(c)). If this point of view is taken, then, based on this work, (i) the traces of superconductivity appear to be observed in SrFe$_2$As$_2$ at very moderate, if not zero, pressure; (ii) it is very conspicuous that $R = 0$ state was not observed in electrical transport measurements under pressure in  SrFe$_2$As$_2$. It is clear that detailed, thermodynamic and transport measurements are needed to address the nature of the $T^*$ feature and to confirm the existence of the superconducting dome in SrFe$_2$As$_2$.
\\

{\it To summarize}, in lightly Sn-doped BaFe$_2$As$_2$ and in SrFe$_2$As$_2$ the structural phase transitions is suppressed by application of pressure approximately two times faster than in non-superconducting SmFeAs(O$_{0.95}$F$_{0.05}$) \cite{uh_sm}. A moderate pressure of $\sim 80$ kbar, or less, is expected to suppress the structural phase transitions completely.

For superconducting RFeAs(O$_{1-x}$F$_x$) a variety of pressure dependencies have been reported: initial increase of $T_c$, followed by a maximum and almost decrease with pressure was reported for LaFeAs(O$_{0.89}$F$_{0.11}$) \cite{jap_pre,zoc_P}, non-linear $T_c$ increase (LaFeAs(O$_{0.95}$F$_{0.05}$) \cite{jap_pre}, SmFeAs(O$_{0.87}$F$_{0.13}$) \cite{uh_sm}) or decrease (CeFeAs(O$_{0.88}$F$_{0.12}$) \cite{zoc_P}), as well as close-to-linear pressure dependencies of different signs and values (RFeAs(O$_{1-x}$F$_x$) \cite{uh_sm,tak_jacs}). The negative, rather large, pressure derivative of $T_c$ observed in (Ba$_{0.55}$K$_{0.45}$)Fe$_2$As$_2$ is well within the range of the available data for oxygen-containing RFeAs(O$_{1-x}$F$_x$). It is possible, if the superconducting phase diagram in (Ba$_{1-x}$K$_{x}$)Fe$_2$As$_2$ has a dome-like shape as a function of K-concentration and pressure, \cite{bri} that (Ba$_{0.55}$K$_{0.45}$)Fe$_2$As$_2$ is positioned in the slightly overdoped region. If compared with the literature data \cite{bri} it appears to be no simple, universal chemical/physical pressure scaling for pure and doped BaFe$_2$As$_2$.

Composite $P - T$ phase diagrams for three parent compounds, AFe$_2$As$_2$ (A = Ba, Sr, Ca), appear to be remarkably similar.

The results above suggest several extensions: (i)detailed study of pressure dependencies in K-doped materials with several values of K-concentrations, if these samples can be reproducibly grown in single crystal form; (ii)high field studies in $H_{c2}(T)$ under pressure; (iii) search for a tetragonal to "collapsed" tetragonal structural phase transition line in AFe$_2$As$_2$ (A = Ba and Sr) and (iv) thermodynamic and transport low temperature measurements in SrFe$_2$As$_2$ under pressure to understand the nature of the $T^*$ anomaly and to confirm the pressure induced superconductivity in this material.

\begin{acknowledgments}
Work at the Ames Laboratory was supported by the US Department of Energy - Basic Energy Sciences under Contract No. DE-AC02-07CH11358.  MST gratefully acknowledges support of the National Science Foundation under DMR-0306165 and DMR-0805335. SLB thanks Starik Kozlodoyev for relevant insights.
\end{acknowledgments}

\clearpage

\begin{figure}
\begin{center}
\includegraphics[angle=0,width=150mm]{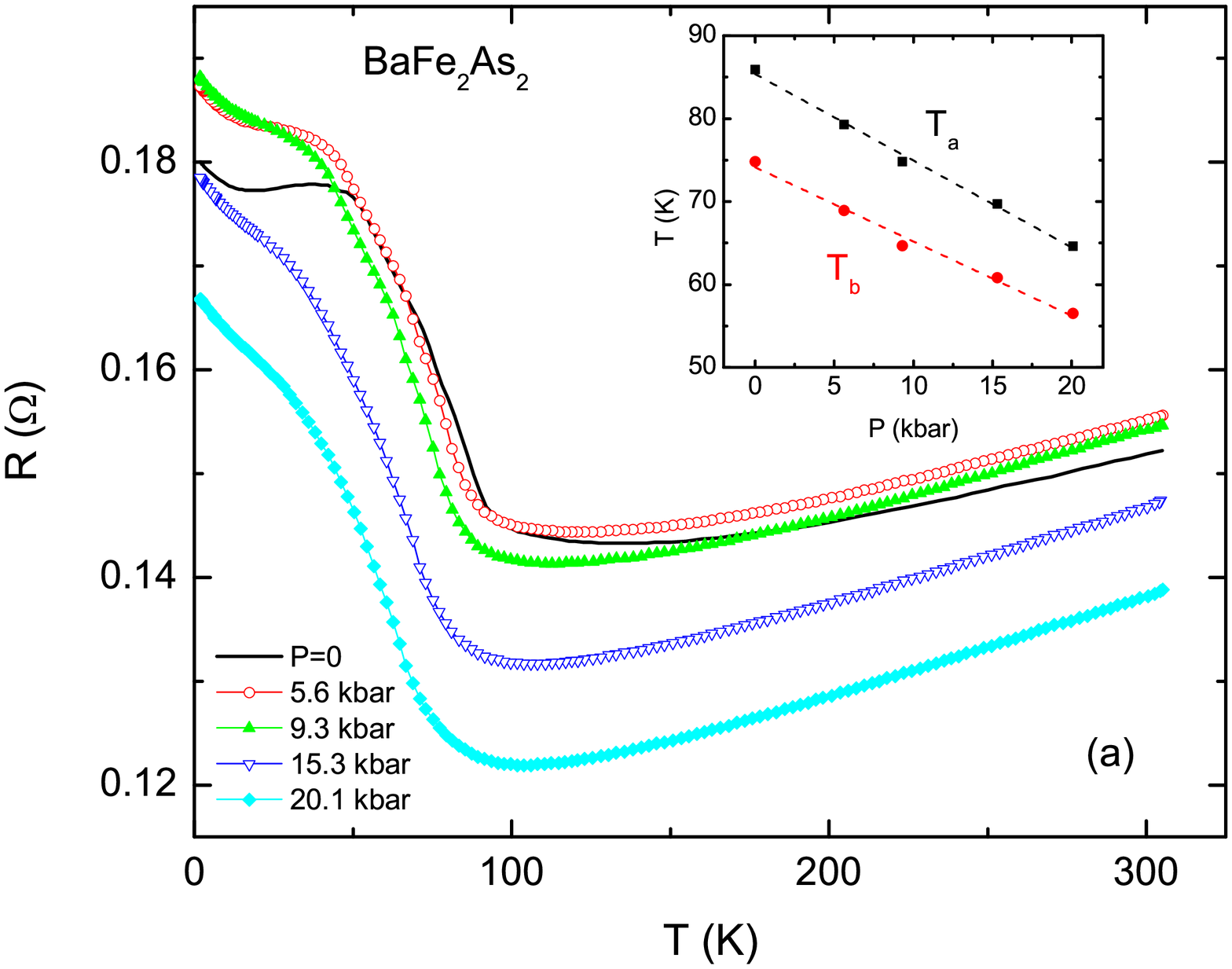}
\end{center}
\end{figure}

\clearpage

\clearpage

\begin{figure}
\begin{center}
\includegraphics[angle=0,width=150mm]{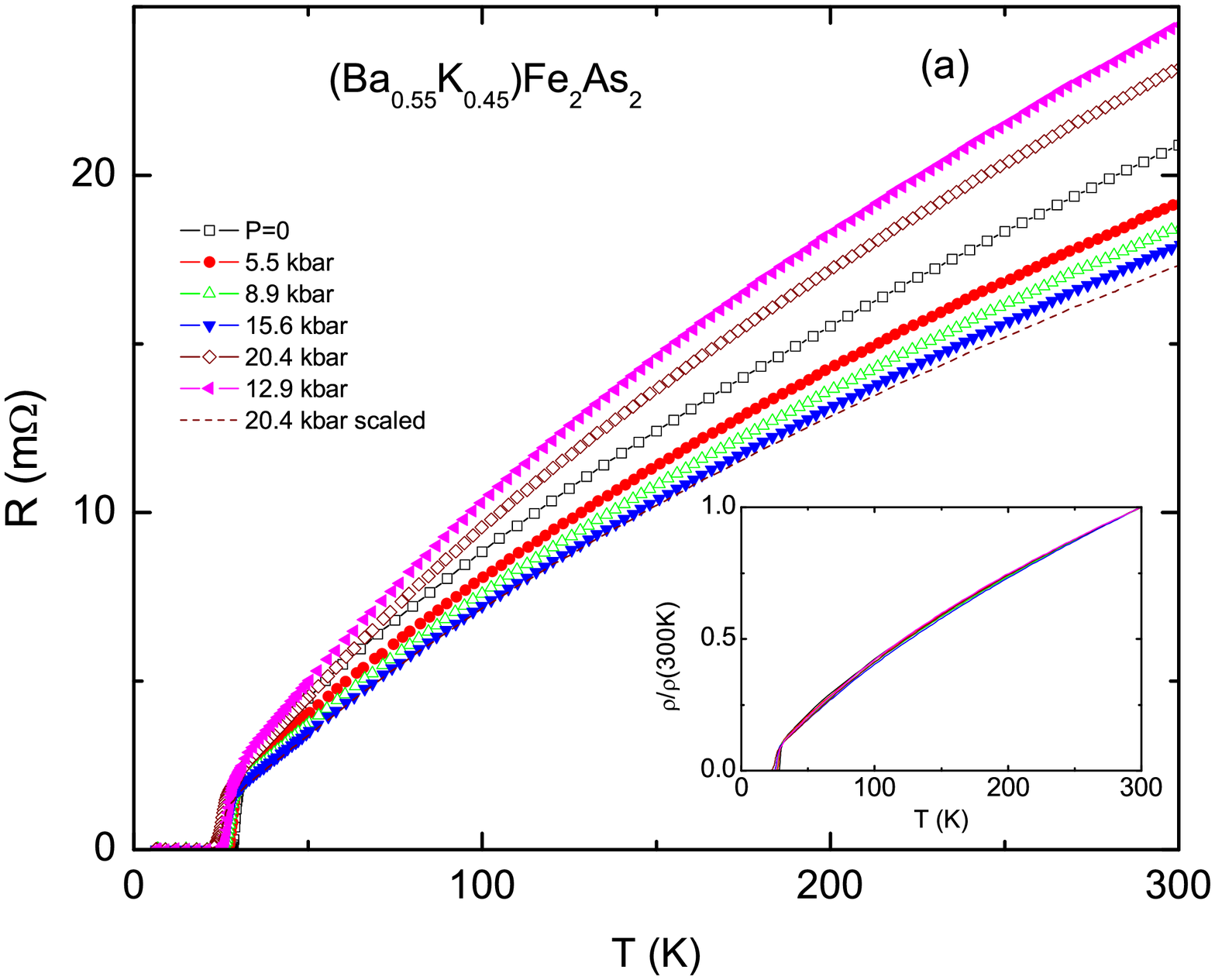}
\end{center}

\end{figure}

\clearpage

\clearpage

\clearpage

\clearpage

\clearpage

\begin{figure}
\begin{center}
\includegraphics[angle=0,width=150mm]{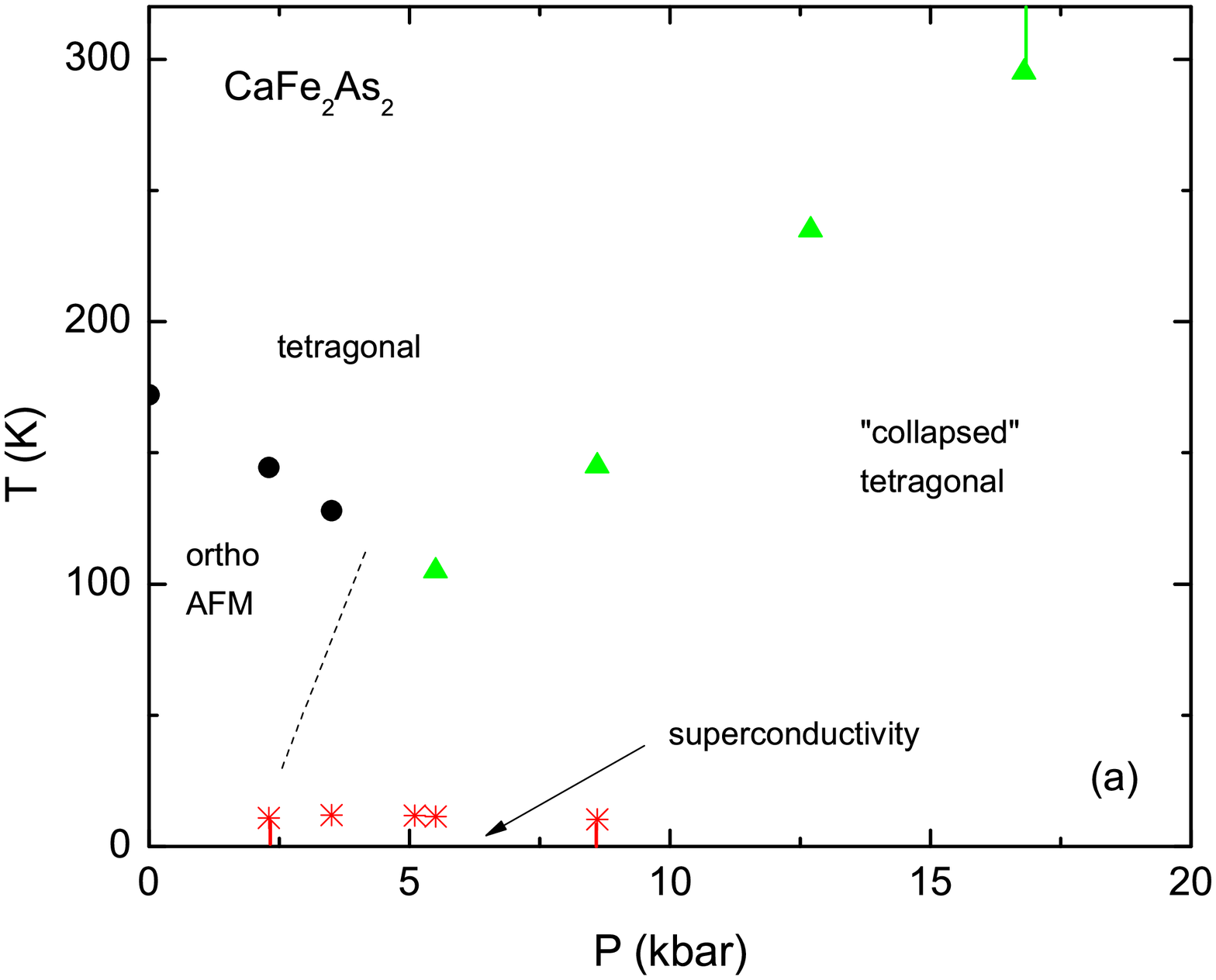}
\end{center}
\end{figure}

\clearpage

\begin{figure}
\begin{center}
\includegraphics[angle=0,width=150mm]{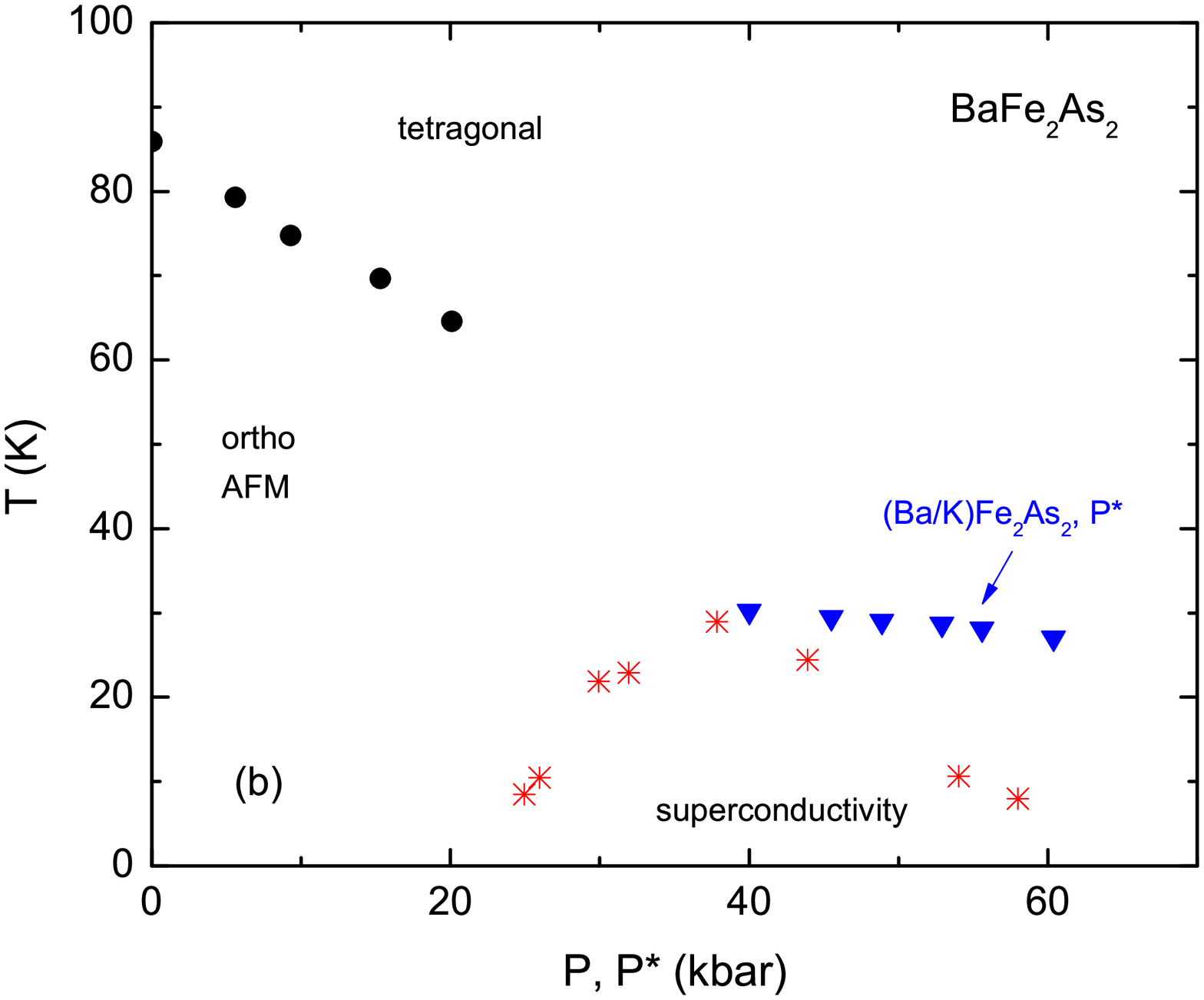}
\end{center}
\end{figure}

\end{document}